\documentclass[%
 reprint,
 amsmath,amssymb,
 aps,
pra,
floatfix,
showkeys
]{revtex4-2}
\usepackage[english]{babel}

\usepackage{amsmath,amsfonts}
\usepackage{textcomp}
\usepackage{stfloats}
\usepackage{url}
\usepackage{verbatim}

\usepackage{graphicx} 
\usepackage[ruled,vlined]{algorithm2e}
\usepackage{braket}
\usepackage{amsmath}
\usepackage{amssymb}
\usepackage{subcaption}
\usepackage{array} 

\usepackage{listings} 
\usepackage{xcolor}

\definecolor{codebg}{RGB}{248,248,248}
\definecolor{codegray}{RGB}{128,128,128}
\definecolor{codepurple}{RGB}{148,0,211}
\definecolor{codeblue}{RGB}{0,102,204}
\definecolor{codegreen}{RGB}{34,139,34}

\lstdefinestyle{mystyle}{
    backgroundcolor=\color{codebg},
    commentstyle=\color{codegreen}\itshape,
    keywordstyle=\color{codeblue}\bfseries,
    numberstyle=\tiny\color{codegray},
    stringstyle=\color{codepurple},
    basicstyle=\ttfamily\small,
    breakatwhitespace=false,
    breaklines=true,
    captionpos=b,
    keepspaces=true,
    numbers=left,
    numbersep=8pt,
    showspaces=false,
    showstringspaces=false,
    showtabs=false,
    frame=single,
    rulecolor=\color{black!30},
    tabsize=4,
    language=Python
}

\begin{document}
\newcolumntype{L}[1]{>{\raggedright\arraybackslash}p{#1}}

\preprint{APS/123-QED}

\title{Blind Transpiler: An open-source library for universally blind and homomorphic quantum computation}

\author{Mohit Joshi}
    \email{joshimohit@bhu.ac.in}
\author{Manoj Kumar Mishra}%
 \email{mkmbhuvns@bhu.ac.in}
\author{S. Karthikeyan}%
 \email{karthik@bhu.ac.in}
\affiliation{%
Department of Computer Science, \\
Banaras Hindu University, Varanasi, India - 221005
}%

\begin{abstract}
Blind quantum computation is a cryptographic primitive that allows a limited-capability client to delegate its complex computation to a remote server without revealing its data and/or computation. This branch of quantum cryptography has been bifurcated into two distinct primitives, quantum homomorphic encryption (concerning the security of only data) and universal blind quantum computation (concerning the security of data and the computing algorithm). These primitives have immense applicability in problems like secure cloud computing, secure quantum variational algorithms, quantum federated learning, and secure multiparty computation. However, no software tools exist for the rapid prototyping of such protocols, hindering the academic interrogation for potential applications. In this paper, we describe the development of the first such library for transpiling circuits written in Qiskit to their blind counterpart, which can then be delegated in a client-server architecture without revealing the client's data and/or computation. The proposed library is designed in modular and reusable component layers, enabling easier scalability to newer BQC primitives and robustness against changes in underlying primitives. We show the implementation of these primitives to a blind variational quantum classifier for the IRIS dataset.
\end{abstract}

\maketitle

\section{Introduction}
Quantum computing threatens the underlying assumption of computational hardness for many classical cryptographic primitives \cite{cain2026shor}. For instance, exponentially fast prime factorization jeopardizes primitives like the RSA algorithm, elliptic curve cryptography, and blockchain \cite{shakib2025impersonation}. Moreover, quadratically faster brute force weakens the hardness of AES, hashing, and digital signatures \cite{ preston2023applying}. 
This poses a grave challenge for today's cryptographic systems. 

On the other hand, quantum computing proposes radically new properties, like the no-cloning theorem, which promises to provide practical information-theoretic security in cryptography. This has been used to strengthen the existing frameworks of security primitives and enables newer solutions unachievable in classical cryptography \cite{fitzsimons_private_2017}.

Blind quantum computation (BQC) is one such primitive that enables a limited-resource client to perform complex quantum computation on a remote server without leaking data and/or computation \cite{fitzsimons_private_2017}. 
These primitives provide solutions to the class of problems that come under `Secure Delegation of Quantum Computation', which is of particular interest for security in today's cloud infrastructure, where quantum computing will be provided as a service over the quantum internet. This is evident from the improving quantum repeater technology \cite{wehner_quantum_2018}. Moreover, many quantum service providers over the classical internet have already started emerging, like IBM, D-Wave, Microsoft, and Rigetti.

The protocols under the BQC primitives work by encrypting the data using classically controlled Pauli's $X$ and $Z$ rotations, which are cheaper and are assumed to be available at the client's side. The outgoing encrypted data using such a technique is proven to be a maximally mixed state and hence independent of the original input data \cite{childs_secure_2005}. The protocols based on this primitive have been shown to provide unconditional security to clients' data and/or computation \cite{broadbent_delegating_2015,broadbent2009universal}. 

The BQC protocols that are concerned with the security of only data provide a secure application of \textit{Quantum Homomorphic encryption} (QHE). This relies on finding an efficient decryption procedure for the delegating gates. Clifford gates can be delegated to the server without any subsequent interaction and corrected at the client side \cite{childs_secure_2005,arrighi_blind_2006}, while the non-Clifford gates require additional corrections delegated to the server side. Several protocols have been proposed for efficient delegation of non-Clifford resources \cite{liang2013symmetric,liang2015quantum,cheng2024quantum,joshi2025quantum}.

Moreover, when the computation is performed over some universal resource set, the technique enables the application of \textit{Universal Blind Quantum Computation} (UBQC). This technique can provide unconditional security to both delegated data and computing algorithms, revealing nothing but the upper limit on the size of data and computation \cite{broadbent2009universal,morimae_blind_2012}. Several universal resource sets have been proposed under different models of computation and resource assumptions \cite{morimae_ground_2011,sueki2013ancilla,zhang_single-server_2018,zhang_hybrid_2019,liu_full-blind_2020,quan2023verifiable,ma_universal_2024,quan2025blind,joshi2026universal}. 

In recent years, numerous applications of these protocols have also been proposed and experimented with.
For instance, protocols for secure variational algorithms using primitives of UBQC and QHE have been presented \cite{shingu2022variational,li2024secure,sulimany2025quantum,sun2024delegated}. 
Several solutions to quantum searchable encryption have also been proposed \cite{fernandez2024implementing,liu2019quantum,joshi2024leveraging}. 
These protocols have been applied to secure quantum approximation algorithms \cite{kim2025secure}, quantum private query \cite{chen2023quantum}, multi-party quantum homomorphic encryption \cite{chen2023practical}, secure Bernstein-Vazirani implementation \cite{fernandez2024homomorphic}, secure fog computing \cite{qu2022secure}, secret sharing protocol \cite{chen2019quantum}, and blind factorization \cite{das2022blind}.
Moreover, several experimental validations for these protocols have also been proposed \cite{fisher2014quantum,tham2020experimental,zeuner2021experimental,gustiani2021blind,li2024experimental}.

Despite the immense applicability, there is no formal software tool for rapid prototyping and academic interrogation of these protocols. This creates a hindrance in ensuring reproducibility and enabling efficient simulated validation of blind protocols. 
In this paper, we fill this research gap by introducing an open-source Python library for converting a quantum circuit written in Qiskit to its blind counterpart, accelerating research in this domain. This software library will enable experimental validation of the existing protocols under various simulation and hardware environments. The software library will also make it easy to generate and validate new applications of BQC protocols for application in both QHE and UBQC.
The main contributions of this paper are:
\begin{itemize}
    \item We present the first software library to perform the transpilation of any quantum circuit to its blind counterpart for rapid prototyping of application frameworks.
    \item The library is capable of transpiling any quantum circuit to its \textit{quantum homomorphic encryption} equivalent, supporting universality using gates from the set $\{H,S,T,CX,CZ,CCX,R_z\}$ as defined in Ref. \cite{childs_secure_2005,broadbent_delegating_2015,tan_universal_2017,fisher2014quantum,joshi2025quantum}.
    \item The library is also capable of transpiling the circuit to its blind counterpart using \textit{universal blind quantum computation}, supporting universality based on the resource set proposed in Ref. \cite{liu_full-blind_2020,zhang_single-server_2018,joshi2026universal}.
    \item The library is designed with modularity and reusability as the primary focus, enabling easier implementation of newer protocols for blind quantum computation using pre-existing transpilation rules or adding newer rules.
    \item The library uses Qiskit primitives and is designed to be robust against internal changes in the Qiskit framework and can be extended to other quantum programming libraries with minimal refactoring.
\end{itemize}

The rest of the paper is structured as follows:
Section \ref{sec:proposed_lib} describes the library, its architecture, and coding style, along with its use case example.
Section \ref{sec:app} provides an illustration of an application of the library for the implementation of a blind variational classifier for the IRIS dataset, and finally, in Section \ref{sec:conclusion}, we provide the concluding remarks.

\begin{figure}[h]
    \centering
    \includegraphics[width=0.7\linewidth]{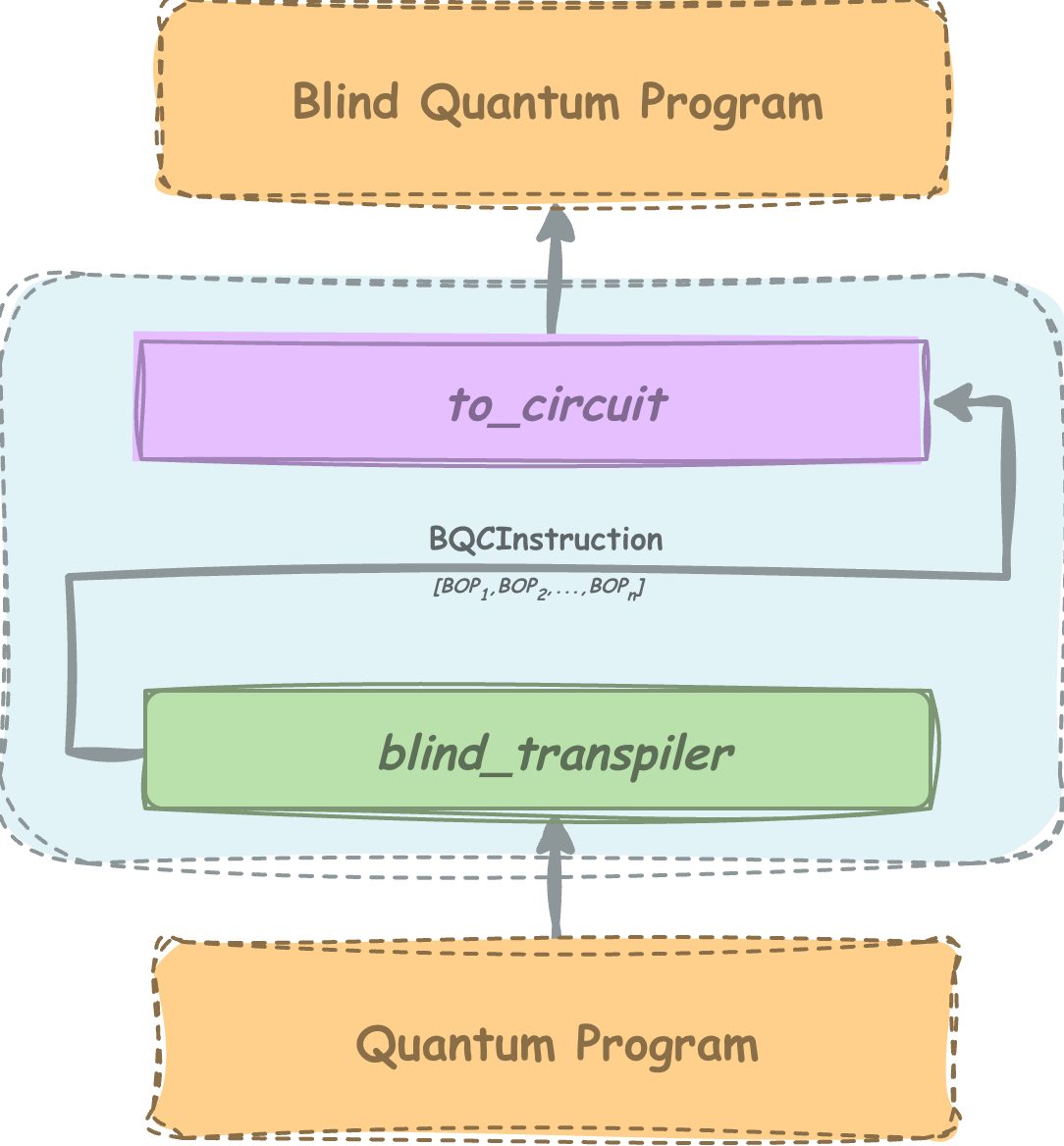}
    \caption{The outline of the \textit{blind\_transpiler} library, which takes a quantum circuit written in Qiskit as input and converts the circuit to an object of the \textit{BQCInstruction} class, which is a collection of modified gate objects \textit{BOP}. This \textit{BQCInstruction} object can then be converted to a blind circuit using the \textit{to\_circuit} function based on the given setting.}
    \label{fig:lib_outline}
\end{figure}

\section{\textit{blind\_transpiler}}\label{sec:proposed_lib}

The proposed \textit{blind\_transpiler} library can be understood as a wrapper around the given quantum circuit, which adds a padding of classically controlled $X$ and $Z$ gates for encryption and their respective decryption gates according to the description of the protocol being implemented. This wrapping essentially enables the secure delegation of the given algorithm with data and/or computation security in a client-server architecture. 
We start by giving a general overview of the functionality and capability of the library in Section \ref{subsec:overview}, which is followed by a detailed discussion of the architecture and design choices made in developement of the library in Section \ref{subsec:lib_arch}, and Section \ref{subsec:code_style} gives an elaboration on usecase of the library with various settings and available options.

\subsection{Overview}\label{subsec:overview}

Fig. \ref{fig:lib_outline} shows the overview of the library functionality.
The library performs this blind transpilation by taking a quantum circuit written in Qiskit as input and generating an object of \textit{BQCInstruction} class as output. This is an internal class of the library which intrinsically encodes a list of modified gate objects, called a blind operation (\textit{BOP}). This \textit{BOP} object contains the definition of the gate along with the additional information about its delegation and classical control.

The \textit{BQCInstruction} object can be directly probed to provide inference about the delegation process, like the actual number of communication rounds and client and server gates needed in the delegation of a given algorithm. This can also be converted to a quantum circuit using the \textit{to\_circuit} function of the \textit{BQCInstruction} class, which can be used for the simulation of the algorithm in a secure delegation environment in various settings (see description, Section \ref{subsec:code_style}).

The library implements the \textit{quantum homomorphic encryption} on the algorithm by defining the translation rules for $H$, $S$, $S^\dagger$, $T$, $T^\dagger$, $CX$, $CZ$, and $CCX$ gates according to Ref. \cite{childs_secure_2005, broadbent_delegating_2015, fisher2014quantum, tan_universal_2017}. Apart from these non-parametric gates, the library can also directly handle the implementation of an arbitrary rotation $R_z$ gate based on the recursive decryption technique as given in Ref. \cite{joshi2025quantum}. This enables the library to perform efficient transpilation of the parametric and non-parametric circuits to their blind counterparts, which is essential for variational circuits with lower depth. 

The library also implements the blind transpilation of a circuit based on \textit{universal blind quantum computation} approaches. This is done by defining the implementation of algorithms over a universal resource set. The library currently implements the blindness of computation based on blind delegation of parametric gates using recursive decryption as described in Ref. \cite{joshi2026universal}. This implementation is most efficient when dealing with circuits having parametric gates. Additionally, two other universal approaches that handle circuits with only non-parametric gates have also been implemented based on an ordered set of universal gates \cite{liu_full-blind_2020} and a universal set based on fixed $\pi/4$ rotation gates \cite{zhang_single-server_2018}.

For all these implementations, the library assumes that the gates from the set $\{X,Z,Swap,Measure\}$ are available with the client, and implementation of all other gates is delegated to the server.

\subsection{Library Architecture}\label{subsec:lib_arch}
The \textit{blind\_transpiler} library is designed to adhere to the principles of modularity, reusability, scalability to other BQC protocols, robustness to changes in Qiskit primitives, and scalability to other quantum programming primitives. 
This is done by dividing the library modules into four layers of abstractions:
\begin{itemize}
    \item \textbf{Layer 0 (Qiskit Primitives):} This layer uses the \textit{QuantumCircuit} and \textit{qiskit.circuit.library} objects from Qiskit library to define native gates.
    \item \textbf{Layer 1 (Core I/O Modules):} This layer provides the core functionality to the library with \textit{orchestrator}, which takes in the input and regulates the process of blind transpilation and \textit{BOP}. This layer interacts with core Qiskit modules to generate gate-like objects, which are packaged as \textit{BQCInstruction} objects and given as the output of the library.
    \item \textbf{Layer 2 (Controllers):} This layer defines different types of algorithms to convert the circuit to its blind counterpart using the rules written in the translator layer. New protocols can be implemented in this layer using predefined translation rules or by adding new translation rules in the translator layer.
    \item \textbf{Layer 3 (Translators):} This layer defines the translation rules that can be used to create the controllers for the blind transpilation. Additional translation rules can be included here for the implementation of new protocols.
\end{itemize}

This layered approach creates a modular design of the library, enabling independence of features and enhancing maintainability of the library. 
The library separates the controllers of blind transpilation and the translation rules, which allows the reusability of rules in different protocols. 
Moreover, the library can also be extended to other protocols by using predefined translation rules or by defining new translation rules. 
Additionally, the interface with Qiskit primitive is limited to only some core I/O modules, which limits the dependence of the library on a particular framework of quantum programming. This enhances the robustness of the library against potential changes in the definition and interfacing of Qiskit primitives and also allows the future extension of the library to other frameworks with minimal refactoring.

\begin{figure*}[h!]
    \centering
    \includegraphics[width=1\linewidth, height=0.35\textheight, keepaspectratio]{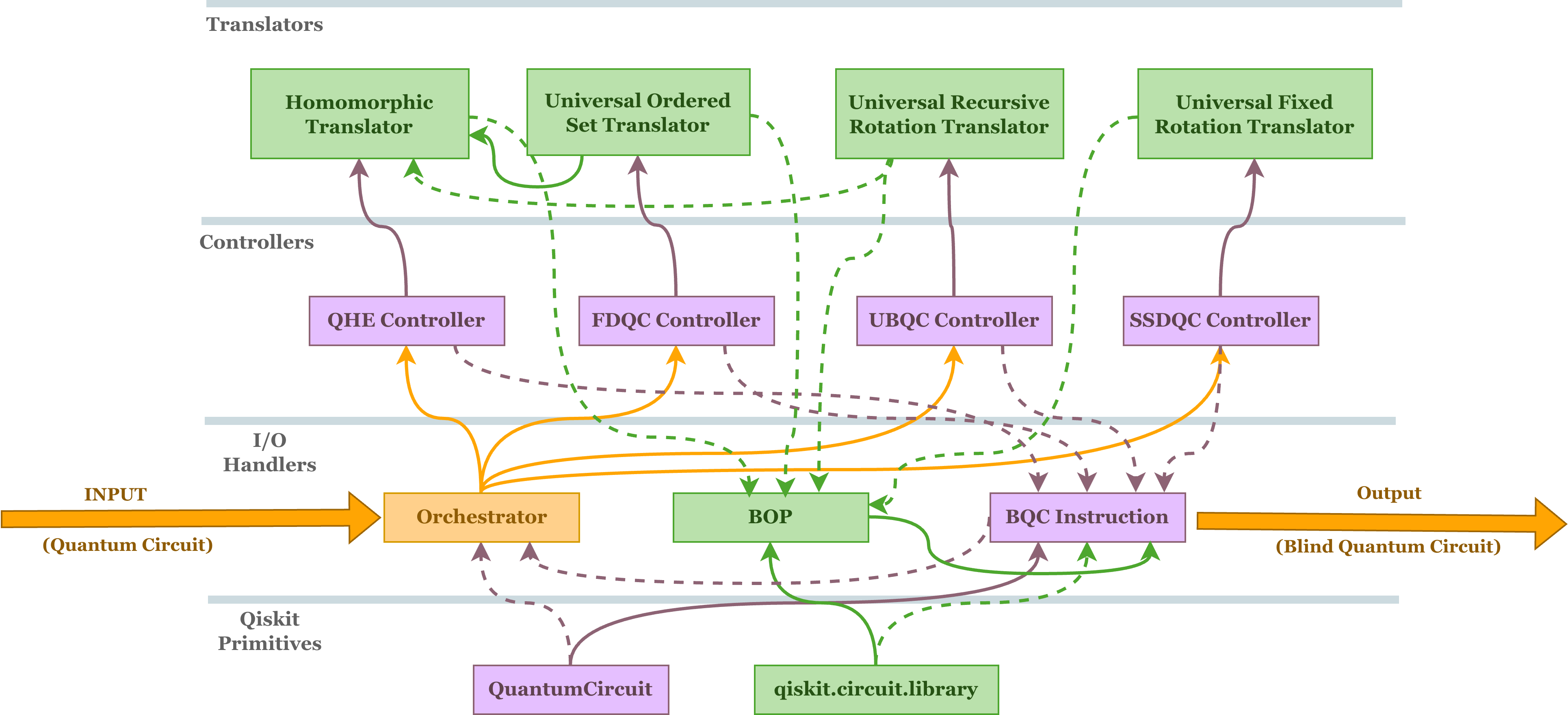}
    \caption{The UML-like relationship between library modules along with their layered architecture. Note that the layered hierarchy enables the library's modularity, reusability, and scalability to new protocols of blindness. The dotted lines show an indirect relationship between the modules and classes. }
    \label{fig:lib_framework}
\end{figure*}

Fig \ref{fig:lib_framework} defines the Unified Modelling Language-like architecture of the library, depicting the major modules in each layer and the relationships among them. Table \ref{tab:function_description} describes the functionality of these modules in more detail. We now describe how these modules enable the blind transpilation of a given circuit.

\begin{table*}
\centering
\setlength{\tabcolsep}{6pt}   
\def\arraystretch{1.5}
\scriptsize
\caption{Summary of the core modules and derived functions of the library that enable the blind transpilation of the given circuit.}
\label{tab:function_description}
\begin{tabular}{| l | l |}
\hline
\multicolumn{2}{|c|}{\small \textbf{Core I/O Modules} \scriptsize} \\
\hline
\parbox{3cm}{\raggedright\textit{BOP}} &
\parbox{10cm}{\raggedright It contains the definition of the gate object to be appended to the circuit, along with the positions of qubits and classical bits. It also contains information regarding the type of operation this gate performs and the condition that defines whether the gate needs to be appended based on the encryption keys and other parameters.} \\
\hline
\parbox{3cm}{\raggedright\textit{BQCInstruction}} &
\parbox{10cm}{\raggedright It contains the list of \textit{BOP} objects and essential information to implement the circuit from the given conditional gates. It also converts the given list of \textit{BOP} gates to a circuit using the \textit{to\_circuit} function, which can generate different types of quantum circuits.} \\
\hline
\parbox{3cm}{\raggedright\textit{BlindTranslator}} &
\parbox{10cm}{\raggedright This module takes the incoming circuit, translates it into available gates in the given format of blind transpilation, and gives the transpiled circuit to the respective controllers. It also has functions to estimate the key size needed for blind transpilation and generate a random key of appropriate size using Python's random function.} \\
\hline
\multicolumn{2}{|c|}{\small \textbf{Controllers for Blind Transpilation} \scriptsize} \\
\hline
\parbox{3cm}{\raggedright\textit{QHE}} &
\parbox{10cm}{\raggedright Converts the circuit to a quantum homomorphic equivalent circuit by iterating over each gate and translating it according to the rules written in the \textit{Homomorphic Translator} library.} \\
\hline
\parbox{3cm}{\raggedright\textit{UBQC}} &
\parbox{10cm}{\raggedright The controller that iterates over the circuit and translates each gate using a universal resource set based on the recursive decryption of $R_z$ gates as defined in the \textit{Universal Recursive Rotation Translator} library.} \\
\hline
\parbox{3cm}{\raggedright\textit{FDQC}} &
\parbox{10cm}{\raggedright Controller that iterates over the given circuit and translates the circuit with the universal resource defined in FDQC with translation rules written in the \textit{Universal Ordered Set Translator} library.} \\
\hline
\parbox{3cm}{\raggedright\textit{SSDQC}} &
\parbox{10cm}{\raggedright The controller that iterates over the given circuit and translates each gate to its $\pi/4$ rotation equivalent as defined in the \textit{Universal Fixed Rotation Translator} library.} \\
\hline
\multicolumn{2}{|c|}{\small \textbf{Modules with Translation Rules} \scriptsize} \\
\hline
\parbox{3cm}{\raggedright\textit{Homomorphic Translator}} &
\parbox{10cm}{\raggedright It contains the translation rules for $H$,$S$,$S^\dagger$,$T$,$T^\dagger$, $CX$,$CZ$, $CCX$ gates by defining the encryption and decryption gates with conditions based on the encryption key and recursive decryption of the $R_z$ gate as defined in Ref \cite{childs_secure_2005, broadbent_delegating_2015,fisher2014quantum, tan_universal_2017, joshi2025quantum}} \\
\hline
\parbox{3cm}{\raggedright\textit{Universal Recursive Rotation Translator}} &
\parbox{10cm}{\raggedright It contains the translation rules for $H$, $R_z$, $CZ$ gate using the universal resource set defined in Ref. \cite{joshi2026universal}} \\
\hline
\parbox{3cm}{\raggedright\textit{Universal Ordered Set Translator}} &
\parbox{10cm}{\raggedright It contains the translation rules for $H$,$S$,$S^\dagger$,$T$,$T^\dagger$, $CX$,$CZ$, $CCX$ using a universal resource defined in Ref. \cite{liu_full-blind_2020}} \\
\hline
\parbox{3cm}{\raggedright\textit{Universal Fixed Rotation Translator}} &
\parbox{10cm}{\raggedright It contains the translation $H$,$S$,$S^\dagger$,$T$,$T^\dagger$, $CX$,$CZ$, $CCX$ using equivalent $\pi/4$ rotations as defined in Ref. \cite{zhang_single-server_2018}} \\
\hline
\end{tabular}
\end{table*}
\normalsize

The library takes in the quantum circuit written using the Qiskit library, which is handled by \textit{orchestrator}. This \textit{orchestrator} is capable of performing the following operations:
    (1) \textit{estimate\_keysize()}, which takes in the circuit along with the format in which blind transpilation has to be done and estimates the size of the key needed for blind transpilation, which can be used with any random number generation technique to generate random keys.   
    (2) \textit{generate\_random\_key()} defines the intrinsic random number generation technique, which uses \textit{key\_size} and \textit{key\_style} variables to generate the random keys using Python's \textit{random} module. 
    (3) \textit{generate\_bqc()} defines the main entry point of the library that takes in the circuit and the format in which the circuit needs to be converted. The circuit is first transpiled according to the basis gates available in the defined format and then given to the controller of the translation library, which outputs the list of modified gate objects \textit{BOP} as an object of the \textit{BQCInstruction} class. The default value of \textit{format} is \textit{qhe}.
    (4) \textit{transpile\_circuit()} defines the Qiskit transpiler, which is used to convert the given circuit to the default basis gates available according to the given format of blindness.

The call of \textit{orchestrator} to different controllers is handled intrinsically by one of four controllers:
    (1) \textit{qhe}, which defines the ancilla resources, client, and server basis needed for quantum homomorphic encryption and operates the protocol that uses translation rules written in \textit{Homomorphic Translator} to convert the circuit to its blind counterpart as defined in Ref. 
    \cite{childs_secure_2005, broadbent_delegating_2015, fisher2014quantum, tan_universal_2017, joshi2025quantum}.
    (2) \textit{ubqc}, which controls the translation of the circuit to a UBQC protocol using \textit{Universal Recursive Rotation Translator} as described in Ref. \cite{joshi2026universal}.
    (3) \textit{fdqc}, which similarly defines the parameters and protocols for translation of a circuit to its blind equivalence using a resource set based on an ordered set of universal gates for UBQC using translation rules written in \textit{Universal Ordered Set Translator} as defined by Ref. \cite{liu_full-blind_2020}.
    (4) \textit{ssdqc}, which similarly defines the transpilation to UBQC defined by rules in \textit{Universal Fixed Rotation Translator} according to Ref. \cite{zhang_single-server_2018}.

These controllers iterate over the given circuit and perform translation by replacing the gate with a list of modified gate objects, called \textit{BOP}, according to the respective translation rule. The \textit{BOP} object contains the definition of the original gate along with its classical control and delegation information. The collection of these \textit{BOP} objects is then packed in a \textit{BQCInstruction} class object, which collectively defines the completed procedure of blind delegation and can be further used with Qiskit primitives to simulate the behaviour of the circuit in a delegated environment. For conversion of the given gate to \textit{BOP}, the library has four predefined translation rule libraries:
    
    (1) \textit{Homomorphic Translator} defines the translation of non-parametric gates from the set $\{H,S,T,CX,CZ,CCX\}$.
    It also defines the translation rule for $S^\dagger(=Z\cdot S)$ and $T^\dagger(=Z\cdot S\cdot T)$ using derived relationship from $S$ and $T$ gates. The translation of the $R_z$ gate is done by first decomposing the given $\theta$ and then performing recursive decryption of subsequent gates.

    (2) \textit{Universal Recursive Rotation Translator} defines the translation rule for universal blind quantum computation using the universal resource set $J(\epsilon)=H_1CZ_{2,3}R_z(\upsilon)_4$, where recursive decryption of arbitrary rotation gates is performed.
    Note that the protocol is optimized with several tweaks in the recursive decryption of parametric gates to make it more efficient in practice.

    (3) \textit{Universal Fixed Rotation Translator} defines the translation rules for $\{H,S,T,CX,CZ,CCX\}$ gates by converting these gates into $\pi/4$ rotation-equivalent gates along with the trap gates.
    The translation rule for $S^\dagger$ and $T^\dagger$ is also implemented using a natural extension of the equivalence of fixed rotation gates. Note that the protocol does not support non-parametric gates; hence, the circuit is pre-transpiled during orchestration to convert parametric gates like $R_z$ to their non-parametric approximation using the \textit{SolovayKitaev} transpilation pass in Qiskit.
        
    (4) \textit{Universal Ordered Set Translator} defines the translation rule using an ordered set of gates from the set $(H,S,CX,CZ,CCX)$ as defined in Ref. \cite{liu_full-blind_2020}. The translation rules from \textit{Homomorphic Translator} are used as the decryption technique
    , with only the difference being the use of a universal resource set. Note that this protocol is also not capable of performing direct decryption of a parametric gate; hence, the circuit is first converted to a non-parametric approximation using the \textit{SolovayKitaev} transpilation pass at the time of orchestration.

In addition to the defined translation rules, custom rules can also be added to the library, which can be used with any predefined controller by changing the default basis gate of the controller.
Moreover, additional controllers can also be added to the library, which can then be listed with the orchestrator to include new functionality in the library.

\subsection{Code Style}\label{subsec:code_style}

\begin{figure}[htbp]
    \centering
    
    \begin{subfigure}{0.45\textwidth}
        \centering
        \includegraphics[width=0.8\linewidth]{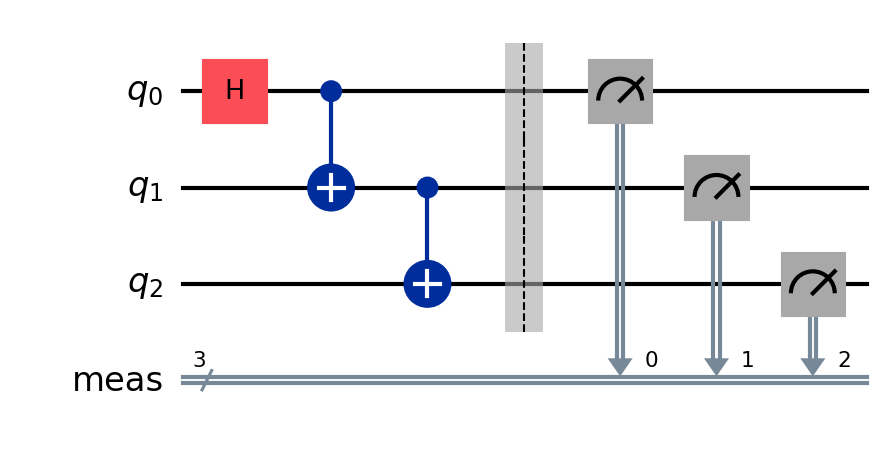}
        \caption{A three-qubit GHZ state circuit}
        \label{fig:original_circuit_3}
    \end{subfigure}
    \hfill
    \begin{subfigure}{0.45\textwidth}
        \centering
        \includegraphics[width=0.6\linewidth]{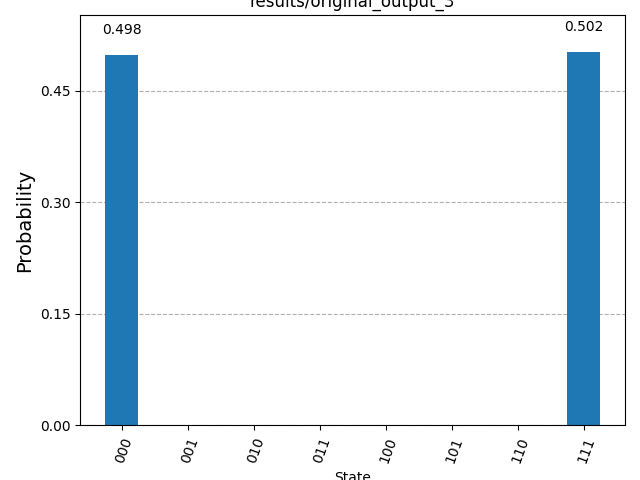}
        \caption{The expected output of the original circuit}
        \label{fig:original_output_3}
    \end{subfigure}
    
    \begin{subfigure}{0.45\textwidth}
        \centering
        \includegraphics[width=0.6\linewidth]{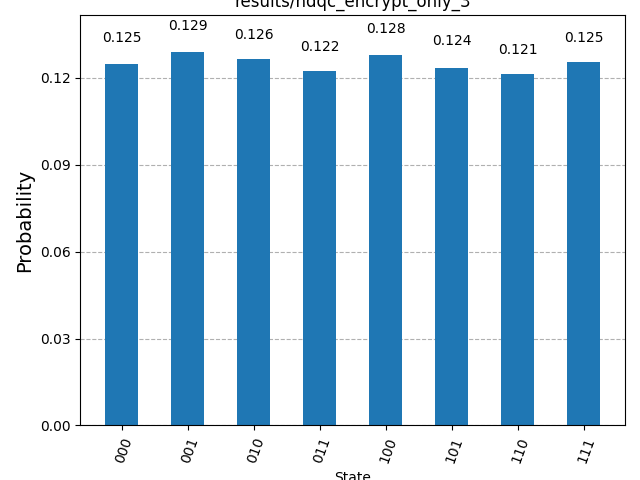}
        \caption{Output without decryption of the circuit sampled over 1000 random keys}
        \label{fig:hdqc_encrypt_only_3}
    \end{subfigure}
    \hfill
    \begin{subfigure}{0.45\textwidth}
        \centering
        \includegraphics[width=0.6\linewidth]{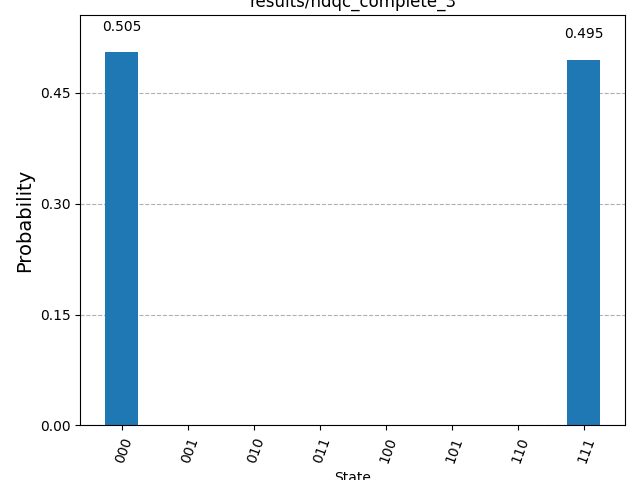}
        \caption{Output with correct decryption procedure sampled over 1000 random keys}
        \label{fig:hdqc_complete_3}
    \end{subfigure}
    \caption{The comparison of the output of the original and the blind transpiled circuit for the circuit that creates the GHZ state. The transpilation has been performed in \textit{qhe} format, and results are shown as averaged over the 1000 randomly sampled keys.}
    \label{fig:circ_3_rslts}
\end{figure}

The \textit{blind\_transpiler} library is written to provide a minimal, clean, and intuitive interface to convert any circuit to an equivalent circuit that provides blindness to data and/or computation, ensures correctness of output, and universality to handle any given circuit. The classes, functions, and variables follow standard and recognizable names. Each class and function is defined with a docstring to elaborate on the functionality of the module.

The transpilation can be easily done using the \textit{generate\_bqc()} function of the library after creating an object of the \textit{BQC} class with minimal parameters as:
\begin{lstlisting}[caption={A simple example of using blind\_transpiler}, label={lst:simple_eg1}]
from blind_transpiler import BQC
bqc = BQC()
bqc_instructions = bqc.generate_bqc(circ=qc, format='qhe')
\end{lstlisting}

Currently, the \textit{blind\_transpiler} supports four different formats of conversion:
\begin{itemize}
    \item 
    \textit{QHE}: This format transpiles a given circuit to its quantum homomorphic encryption-based implementation, which provides security to the client's data as described in Ref. \cite{childs_secure_2005, broadbent_delegating_2015,fisher2014quantum, tan_universal_2017, joshi2025quantum}.
    \item 
    \textit{UBQC}: This format transpiles the given circuit to a universal blind counterpart using a universal resource set based on recursive decryption of rotation gates. The protocol provides security to both data and computation and can natively handle the parametric gates. The format performs blind decryption as described in Ref. \cite{joshi2026universal}.
    \item 
    \textit{FDQC}: This format transpiles the given circuit to its universal blind equivalent, where both data and computation will be secure with delegation. This approach is based on Ref. \cite{liu_full-blind_2020} and handles parametric gates by approximating them using the Solovay-Kitaev decomposition, as the approach cannot handle parametric gates natively.
    \item 
    \textit{SSDQC}: This format transpiles the quantum circuit to its fixed-rotation equivalent universal blind counterpart. This protocol also provides security to both data and computation as described in Ref. \cite{zhang_single-server_2018}. It also handles the parametric gates by approximating them using the Solovay-Kitaev decomposition, as the protocol cannot decrypt parametric gates natively.
\end{itemize}

The simplest implementation given in Listing \ref{lst:simple_eg1} intrinsically assumes a key of appropriate size, which can be probed using the \textit{bqc.key} variable. However, we can also define the random key explicitly by first estimating the key size using \textit{estimate\_keysize} function and then using any technique to generate a list of binary random numbers, or we can also use the \textit{generate\_random\_key} function of the library that generates the random key using Python's \textit{random} function as given in Listing \ref{lst:eg_with_estimate_keysize}.
\begin{lstlisting}[caption={Example with explicitly defined size of key}, label={lst:eg_with_estimate_keysize}]
key_size = bqc.estimate_keysize(format='qhe', circ=qc)
keys =  bqc.generate_random_key(key_size=key_size, key_style='rand')
bqc_instructions  = bqc.generate_bqc(circ=circ, format ='qhe', keys=keys)
\end{lstlisting}
The keys can be generated in four random styles, namely, \textit{all\_0}, \textit{all\_1}, \textit{rand\_fix}, \textit{rand}. 

The generated output, \textit{bqc\_instruction}, is an object of class \textit{BQCInstruction}, which can be further probed to analyze the static behaviour of the algorithm after transpilation. This includes information regarding the number of client and server gates, additional gates used in encryption, 
and the communication rounds needed for the protocol to run, as shown in Listing \ref{lst:probing_bqc_instruction}.
\begin{lstlisting}[caption={Probing the object of BQCInstruction class}, label={lst:probing_bqc_instruction}]
print(bqc_instruction.n_client_gates)
print(bqc_instruction.n_server_gates)
print(bqc_instruction.n_secure_gates)
print(bqc_instruction.n_communication_rounds)
\end{lstlisting}
Moreover, the \textit{bqc\_instruction} is a collection of \textit{BOP} objects, which are modified gate objects that can be iterated over as a list for further manipulation, as shown in Listing \ref{lst:probing_bop}.
\begin{lstlisting}[caption={Probing the object of BOP class}, label={lst:probing_bop}]
for elem in bqc_instructions:
    print(elem.op_type)
    print(elem.conditional)
    print(elem.gate)
\end{lstlisting}
Here, the \textit{op\_type} contains the information about the type of operation, namely, client-implementable, encryption gate, decryption gate, or server-implementable gates. This operation type instructs the protocol to decide if the gate needs to be delegated to the server or if the client is able to perform it.  The \textit{conditional} variables define whether the gate will be needed in the final circuit based on the encryption key and decryption process of the gates involved.

For simulation purpose, the \textit{to\_circuit} function of \textit{bqc\_instruction} object can be used as shown in Listing \ref{lst:create_qc_from_bqc}:
\begin{lstlisting}[caption={Creating quantum circuit from BQCInstruction object}, label={lst:create_qc_from_bqc}]
blind_circ = bqc_instructions.to_circuit(const_type='complete', show_barrier=True) 
\end{lstlisting}
This gives us a circuit written in Qiskit. This circuit can then be simulated using Qiskit Aer or processed on real hardware.
The \textit{const\_type} variable assists in the creation of the circuit, which can be one of four types:
\begin{itemize}
    \item \textit{complete}: creates the complete circuit with client-implementable, server-implementable, and security gates.
    \item \textit{client\_only}: creates the circuit which includes only gates implemented at the client side, that are client-implementable gates and security gates.
    \item \textit{server\_only}: creates the circuit that contains only server-implementable gates.
    \item \textit{encrypt\_only}: creates the circuit where all the decryption gates are omitted from the circuit and essentially demonstrates the view at the server side without correct decryption.
\end{itemize}

We now use the library to perform blind delegation of some simple circuits and showcase the correctness and blindness of the converted circuit.
As the possible key space is generally very large, we have performed random sampling of output over 1000 random keys. 

Fig. \ref{fig:original_circuit_3} shows the circuit diagram for a Greenberger–Horne–Zeilinger (GHZ) state, which has an expected output as shown in Fig. \ref{fig:original_output_3}. 
The results without decryption and with correct decryption are shown in Fig. \ref{fig:hdqc_encrypt_only_3} and Fig. \ref{fig:hdqc_complete_3}, respectively. The results show that over a randomly sampled circuit, the output will always be in a maximally mixed state if the decryption has not been performed properly. 

\section{Applications to Blind Quantum Machine Learning}\label{sec:app}

\begin{figure}
    \centering
    
    \begin{subfigure}{0.35\textwidth}
        \centering
        \includegraphics[width=0.9\linewidth]{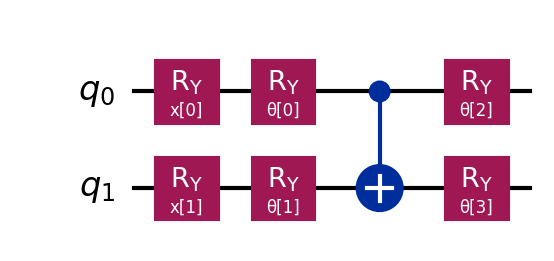}
        \caption{Variational Circuit used for training on the IRIS dataset}
        \label{fig:original_circuit_iris}
    \end{subfigure}
    \hfill
    \begin{subfigure}{0.35\textwidth}
        \centering
        \includegraphics[width=0.9\linewidth]{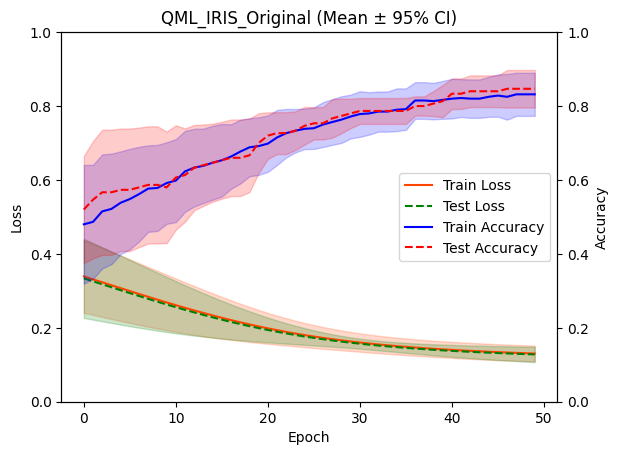}
        \caption{The expected output of the original circuit}
        \label{fig:original_output_iris}
    \end{subfigure}
    
    \begin{subfigure}{0.35\textwidth}
        \centering
        \includegraphics[width=0.9\linewidth]{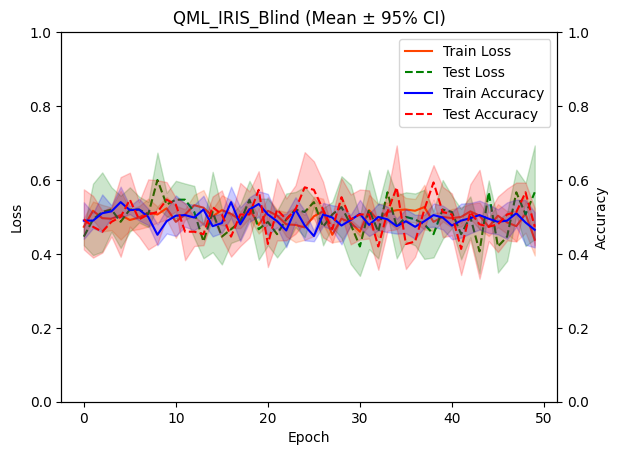}
        \caption{Output without decryption of the circuit sampled over 1000 random keys}
        \label{fig:hdqc_encrypt_iris}
    \end{subfigure}
    \hfill
    \begin{subfigure}{0.35\textwidth}
        \centering
        \includegraphics[width=0.9\linewidth]{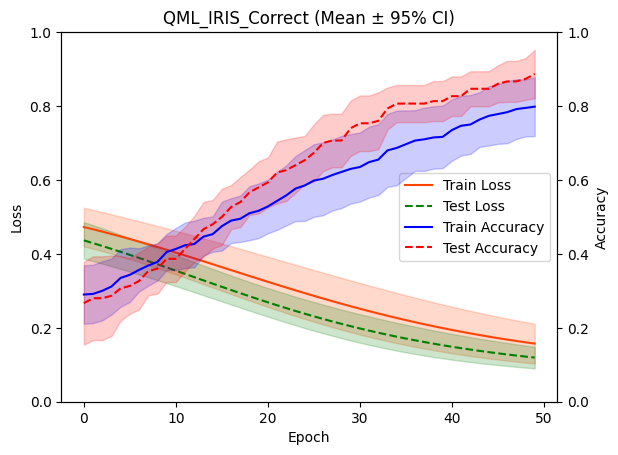}
        \caption{Output with correct decryption procedure sampled over 1000 random keys}
        \label{fig:hdqc_complete_iris}
    \end{subfigure}
    \caption{The comparison of the output of the original and the blind transpiled circuit. (a) shows the original parametric circuit used, (b) shows its desired output, (c) shows the output if the decryption is not performed, and (d) shows the output if the correct decryption has been performed. Note that the transpilation has been performed in \textit{qhe} format, and results are shown as averaged over the 1000 randomly sampled keys.}
    \label{fig:circ_iris_rslts}
\end{figure}

The most vital near-term application of quantum computing is supposed to be through hybrid quantum-classical algorithms using variational quantum circuits. It has been proposed that a quantum circuit can learn faster with fewer parameters and often overcome the barren plateau problem. Hence, a secure alternative to cloud computing will have variational algorithms at the center. Many techniques to provide secure quantum variational learning algorithms have been proposed using blind quantum computing primitives \cite{shingu2022variational,li2024secure,sulimany2025quantum}.

Here, we perform an experiment to train a parametric circuit for classification on the IRIS dataset. The dataset contains 150 data points of iris samples with three classes: Setosa, Versicolor, and Virginica.

For simplicity of the classification task, we have only two features, namely, sepal length and sepal width, and the problem is simplified by considering binary classification, including only the presence and absence of the Setosa class specimen. The data values have been standardized using \textit{StandardScaler}.

\begin{lstlisting}
from sklearn.datasets import load_iris
from sklearn.preprocessing import StandardScaler

iris = load_iris()
X = iris.data[:, :2]   # use only 2 features for simplicity
y = iris.target

# Binary classification (class 0 vs rest)
y = (y == 0).astype(int)

scaler = StandardScaler()
X = scaler.fit_transform(X)

\end{lstlisting}

For training, a two-qubit variational circuit with four parameters is considered, as shown in Fig. \ref{fig:original_circuit_iris}. 

\begin{lstlisting}
from qiskit import QuantumCircuit
from qiskit.circuit import ParameterVector
from qiskit.primitives import StatevectorEstimator
from qiskit.quantum_info import Pauli

n_qubits = 2
params = ParameterVector('theta', length=4)
observable = Pauli("ZI")
estimator = StatevectorEstimator()

def create_circuit(x):
    qc = QuantumCircuit(n_qubits)
    
    # Data encoding
    for i in range(n_qubits):
        qc.ry(x[i], i)
    
    # Variational layer
    qc.ry(params[0], 0)
    qc.ry(params[1], 1)
    qc.cx(0, 1)
    qc.ry(params[2], 0)
    qc.ry(params[3], 1)
    
    return qc
\end{lstlisting}

The blind transpilation is performed using the following function:

\begin{lstlisting}
from blind_transpiler import BQC
  
def blind(circ, blind_format='qhe', const_type='complete', key_style ='rand', keys=None):
    bqc = BQC()
    if keys == None: keys =  bqc.generate_random_key(format=blind_format,  circ=circ, key_style=key_style)
    bqc_instructions  = bqc.generate_bqc(circ=circ, format =blind_format, keys=keys)
    blind_circ = bqc_instructions.to_circuit(const_type=const_type, show_barrier=True)
    return blind_circ

\end{lstlisting}

The training is performed over $50$ epochs, and statistical variance over $5$ different runs of the experiment has been considered. A train-test split of $20\%$ has been considered with a learning rate of $0.1$.

Fig. \ref{fig:original_output_iris} shows the expected training and testing trends of the simulation for $50$ epochs averaged over $5$ runs. Fig. \ref{fig:hdqc_encrypt_iris} shows the result without proper decryption, which shows that nothing of value is learned while the decryption is not performed properly. The results in Fig. \ref{fig:hdqc_complete_iris} show the outcome with correct decryption averaged over $5$ experiment runs. The outcome shows the correct learning pattern when the decryption has been performed properly. 

\section{Conclusion}\label{sec:conclusion}
Blind quantum computation is a primitive of secure computation which allows a limited-resource client to delegate its complex computation to a remote server without revealing data and/or the algorithm. Many techniques to perform BQC have been proposed using various models of computation. Moreover, this paradigm has proven application in varied fields, including quantum searchable encryption, secure quantum machine learning, quantum homomorphic encryption, and secure multi-party quantum computation. However, the lack of software tools for rapid prototyping of applications of such protocols creates a hindrance in academic endeavours. 
We develop the first library capable of transpiling a given circuit to its blind counterpart, enabling rapid testing of applications and frameworks designed using such paradigms of security. The library is designed with multiple layers of abstraction, allowing the design to be modular, reusable, and scalable. Right now, the library is capable of implementing four variants of circuit-based blind quantum computation as described in Ref. \cite{joshi2025quantum,joshi2026universal,liu_full-blind_2020,zhang_single-server_2018}. However, other protocols can be easily integrated into the library by expanding the controllers using existing or new translation rules with minimal refactoring in the code. The interface with the Qiskit primitive has also been kept to a minimum, allowing for expansion of the library to include quantum circuits written in other languages. 
Moreover, other properties like verification and authentication can also be considered in the future extension of library protocols.

\section*{Software Availability.}

BlindTranspiler is an open-source Python library for blind quantum computation and quantum homomorphic encryption.
The software is archived on Zenodo (\url{https://doi.org/10.5281/zenodo.21437379}) \cite{joshi2026blindtranspiler},
ensuring long-term availability and reproducibility of the results reported in this work.
The package is also distributed through PyPI and can be installed using
\begin{verbatim}
pip install blind-transpiler
\end{verbatim}
The source code is maintained on GitHub at
\url{https://github.com/joshiCoding/blind_transpiler} and for complete documentation visit \url{https://blindtranspiler.joshicoding.in/}.

\section*{Acknowledgments}
The authors acknowledge the use of IBM Quantum services for this work. The views expressed are those of the authors and do not reflect the official policy or position of IBM or the IBM Quantum team. 
The authors also acknowledge the National Supercomputing Mission (NSM) for providing computing resources of ‘PARAM Shivay’ at the Indian Institute of Technology (BHU), Varanasi, which is implemented by C-DAC and supported by the Ministry of Electronics and Information Technology (MeitY) and the Department of Science and Technology (DST), Government of India. 
This research is supported by a seed grant under the IoE, BHU [grant no. R/Dev/D/IoE/SEED GRANT/2020-21/Scheme No. 6031].

\bibliography{main_ref}

\end{document}